\documentclass[3p,times,twocolumn]{elsarticle}

\usepackage{ecrc}

\volume{00}

\firstpage{1}

\journalname{Nuclear Physics B Proceedings Supplement}

\runauth{}

\jid{nuphbp}

\jnltitlelogo{Nuclear Physics B Proceedings Supplement}

\usepackage{amssymb,amsmath}

\usepackage[figuresright]{rotating}

\begin{document}

\begin{frontmatter}



\dochead{}

\title{Transition form factors of $\pi^0$, $\eta$ and
$\eta'$ mesons: \\ What can be learned from
anomaly sum rule?}


\author[JINR]{Yaroslav~Klopot\corref{cor1}\fnref{fn1}}
\ead{klopot@theor.jinr.ru}
\fntext[fn1]{On leave from  Bogolyubov Institute for Theoretical Physics, 03680 Kiev, Ukraine}
\author[JINR,ITEP]{Armen Oganesian}
\ead{armen@itep.ru}
\author[JINR]{Oleg Teryaev}
\ead{teryaev@theor.jinr.ru}

\address[JINR]{Bogoliubov Laboratory of Theoretical Physics, Joint Institute for Nuclear Research, 141980 Dubna, Russia}
\address[ITEP]{Institute of Theoretical and Experimental Physics, 117218  Moscow, Russia}

\begin{abstract}
We have studied the applications of the anomaly sum rule to the transition form factors of light pseudoscalar mesons: $\pi^0$, $\eta$ and $\eta'$. This nonperturbative QCD approach can be used even if the QCD factorization is broken. The possibility of small non-OPE corrections to the spectral density in view of the recent BaBar and Belle data is explored. The accuracy of the applied method and the role of the quark mass contributions are discussed.
\end{abstract}


\end{frontmatter}


\section{Introduction}
The recent progress in experimental study of the transition form factors (TFFs) of pseudoscalar mesons $F_{P\gamma}$ (P=$\pi^0$, $\eta$,  $\eta'$) \cite{Aubert:2009mc,Uehara:2012ag,BABAR:2011ad} have drawn a significant attention as it challenged our theoretical understanding of the processes $\gamma \gamma^* \to P$ at high spacelike momentum transfer $Q$. 
The measurements of the pion TFF were performed by \textit{BaBar} \cite{Aubert:2009mc} and \textit{Belle} \cite{Uehara:2012ag} collaborations up to $Q^2=35$ GeV$^2$ and appear to be quite inconsistent. To top it all, the \textit{BaBar} measurements at $Q^2>10$ GeV$^2$ reveal a certain excess of the pion TFF over the theoretically predicted (from pQCD)  asymptote \cite{Lepage:1980fj}  and cannot be satisfactory explained within the QCD factorization approach \cite{Stefanis:2012yw,Agaev:2010aq}.

In this work we demonstrate the applications of the axial anomaly in its dispersive representation to the  pion \cite{Klopot:2010ke,Klopot:2012hd} as well as to the $\eta$ and  $\eta'$ \cite{Klopot:2012hd,Klopot:2011qq,Klopot:2011ai} meson TFFs without relying on the QCD factorization. We pay also a special attention to the quark mass effects in the explored relations for the TFFs.

\section{Anomaly sum rule and hadron contributions}

The axial anomaly \cite{Bell:1969ts} plays an important role not only in the case of real photons (where, e.g., it is known to govern the neutral pion decay $\pi^0 \to \gamma\gamma$) but also in the case of virtual photons. The dispersive approach to axial anomaly  \cite{Dolgov:1971ri} leads to the anomaly sum rule (ASR) \cite{Horejsi:1985qu,Horejsi:1994aj},  which can be related to $\pi^0$, $\eta$ and $\eta'$ TFFs  and gives a tool to explore the TFFs even beyond the QCD factorization \cite{Klopot:2010ke,Klopot:2012hd,Klopot:2011qq,Klopot:2011ai,Melikhov:2012qp}.

Let us briefly remind the derivation of the ASR. The VVA triangle graph amplitude contains an axial current and two vector currents 

\begin{equation} \label{VVA} 
T_{\alpha \mu\nu}(k,q)=\int
d^4 x d^4 y e^{(ikx+iqy)} \langle 0|T\{ J_{\alpha 5}(0) J_\mu (x)
J_\nu(y) \}|0\rangle, 
\end{equation}
where $k$ and $q$ are the photons' momenta. This amplitude can be presented as a tensor decomposition  

\begin{align}
\label{eq1} \nonumber T_{\alpha \mu \nu} (k,q)  & =  F_{1} \;
\varepsilon_{\alpha \mu \nu \rho} k^{\rho} + F_{2} \;
\varepsilon_{\alpha \mu \nu \rho} q^{\rho}
\\ \nonumber
  & + \; \; F_{3} \; k_{\nu} \varepsilon_{\alpha \mu \rho \sigma}
k^{\rho} q^{\sigma} + F_{4} \; q_{\nu} \varepsilon_{\alpha \mu
\rho \sigma} k^{\rho}
q^{\sigma}\\
  & + \; \; F_{5} \; k_{\mu} \varepsilon_{\alpha \nu
\rho \sigma} k^{\rho} q^{\sigma} + F_{6} \; q_{\mu}
\varepsilon_{\alpha \nu \rho \sigma} k^{\rho} q^{\sigma},
\end{align}
where   $F_{j} = F_{j}(p^{2},k^{2},q^{2}; m^{2})$, $j = 1, \dots ,6$, $p=k+q$, are the Lorentz invariant amplitudes.

The ASR can be obtained from consideration of the unsubtracted   dispersion relations which result in the finite subtraction for axial current divergence. In the case of isovector $J^{(3)}_{\alpha 5}=  \frac{1}{\sqrt{2}}(\bar{u} \gamma_{\alpha} \gamma_5 u - \bar{d} \gamma_{\alpha} \gamma_5 d)$ and octet $J_{\alpha 5}^{(8)}=\frac{1}{\sqrt{6}}(\bar{u} \gamma_{\alpha} \gamma_5 u + \bar{d} \gamma_{\alpha} \gamma_5 d - 2 \bar{s} \gamma_{\alpha}\gamma_5 s)$ currents for  the kinematical configuration we are interested in ($k^2=0$, $-q^2=Q^2\ge 0$) the  ASR reads \cite{Horejsi:1994aj}  

\begin{equation}\label{asr}
\int_{0}^{\infty} A_{3}^{(a)}(s,Q^{2}; m_i^{2}) ds =
\frac{1}{2\pi}N_c C^{(a)} \;, a=3,8,
\end{equation}
where $A_{3} = \frac{1}{2}Im_{p^2} (F_3-F_6)$, $N_c=3$ is a number of colors, $C^{(3)}=\frac{1}{3\sqrt{2}}$ and $C^{(8)}=\frac{1}{3\sqrt{6}}$ are charge factors, and $m_i$ are quark masses. These ASR relations hold for any $Q^2$ and any quark masses $m_i$. They are also "exact" relations: $\alpha_s$ corrections are zero and it is expected that all nonperturbative corrections are absent as well (due to 't Hooft's principle \cite{'tHooft:1980xb,Horejsi:1994aj}).  

Saturating the lhs of the three-point correlation function (\ref{VVA}) with the resonances and singling out their contributions to the ASR (\ref{asr}), we get a sum of resonances $M$ with appropriate quantum numbers

\begin{equation} \label{qhd}
\pi \sum f_M^a F_{M\gamma} = \int_{0}^{\infty} A_{3}^{(a)}(s,Q^{2};m^{2})
ds, 
\end{equation}
where the decay constants $f_M^a$ and TFFs $F_{M\gamma}$ are defined as follows,
\begin{align} \label{def_f}
\langle& 0|J^{(a)}_{\alpha 5}(0) |M(p)\rangle=
i p_\alpha f^a_M, \\
\int & d^{4}x e^{ikx} \langle M(p)|T\{J_\mu (x) J_\nu(0)
\}|0\rangle=\epsilon_{\mu\nu\rho\sigma}k^\rho q^\sigma
F_{M\gamma}.
\end{align}

\section{Isovector channel of ASR and $\pi^0$ TFF}
Singling out the first (pion) contribution in (\ref{qhd}) and introducing the continuum threshold in the isovector channel $s_3$, the ASR (\ref{asr}) leads to
\begin{equation} \label{qhd3}
\pi f_{\pi}F_{\pi\gamma}(Q^2)+ \int_{s_3}^{\infty} A_{3}^{(3)}(s,Q^{2}) ds  =\frac{1}{2\pi}N_c C^{(3)}.
\end{equation}
For a given flavor $q$,  the one-loop approximation for the spectral density is
\begin{align} \label{a3}
A_{3}^{(q)}(s,Q^{2})=\frac{e_q^2N_c }{2\pi}\frac{\Theta (s-4m_q^2)}{(Q^2+s)^2}  \left(Q^2 R^{(q)} +2m_q^2\ln\frac{1+R^{(q)}}{1-R^{(q)}}\right) ,
\end{align}
where $R^{(q)}=\sqrt{1-4m_q^2/s}$, $e_q$ are quark charges in the units of the absolute value of electron charge. \\
Then, as $A_3^{(3)}=\frac{1}{\sqrt{2}}(A_3^{(u)}-A_3^{(d)})$, the ASR (\ref{qhd3}) yields \cite{Klopot:2012hd} the pion TFF  ($m_u=m_d=m$),
\begin{align} \label{f3m}
F_{\pi\gamma}(Q^2;m^2)=\frac{1}{2\sqrt{2}\pi^2f_{\pi}}\frac{s_3}{s_3+Q^2}\times \nonumber\\ \Bigl [1-\frac{2m^2}{s_3}(\frac{2}{R_3+1}+\ln\frac{1+R_3}{1-R_3})\Bigr ],
\end{align}
where  $R_3=\sqrt{1-4m^2/s_3}$. This equation is obtained without QCD factorization hypothesis and is valid at any $Q^2$, including $Q^2=0$. At the same time, supposing that QCD factorization is valid at $Q^2 \to \infty$ and employing the pQCD predicted  asymptotic value for the pion TFF \cite{Lepage:1980fj} $Q^2F_{\pi\gamma}^{as}(Q^2)=\sqrt{2}f_\pi$, $f_\pi=130.4$ MeV the continuum threshold  $s_3$ can be determined from (\ref{f3m}), $s_3=4\pi^2f_\pi^2$ GeV$^2$. Then, if we put $m=0$, Eq. (\ref{f3m}) reproduces a well-known Brodsky-Lepage interpolation formula \cite{Brodsky:1981rp} (it was also proved in a different approach in \cite{Radyushkin:1995pj}).
Numerically, $s_3=0.67$ GeV$^2$ appears to be rather close to the continuum threshold extracted from the two-point QCD sum rules.

The obtained expression for the pion TFF (\ref{f3m}) has no
$\alpha_s$ corrections \cite{Jegerlehner:2005fs}; the higher $\alpha_s$ correction are not excluded, but they, as well as the possible higher power OPE corrections (e.g., quark and gluon condensates) should be numerically small. 
The $u$- and $d$-quark mass contributions in (\ref{f3m}) appear to be numerically negligible ($\sim 0.15\%$ of the main term). The quark mass corrections, however, are not so small in the octet channel, where $s$-quark contributes. We will discuss this issue in the next section.

Let us note, that  the continuum contribution vanishes in the case the chiral limit and $Q^2=0$,  $\int_{s_3}^{\infty} A_{3}^{(3)}(s,Q^{2}) ds\propto\frac{Q^2}{Q^2+s_3}\to 0$ at $Q^2\to 0$. This conforms with the fact, that the contributions of the pseudovector and higher pseudoscalar (except $\pi^0$) mesons (in Eq. (\ref{qhd3})) vanish in this case: the first ones do not give contributions as they do not decay into 2$\gamma$, while the second ones' contributions vanish as their decay constants (4) vanish due to conservation of the axial current in the chiral limit (up to electromagnetic corrections), $f_{M}^{(3)}\propto m_\pi^2/m_{M}^2 \to 0$. 

\begin{figure}
\centerline{
\includegraphics[width=0.5\textwidth]{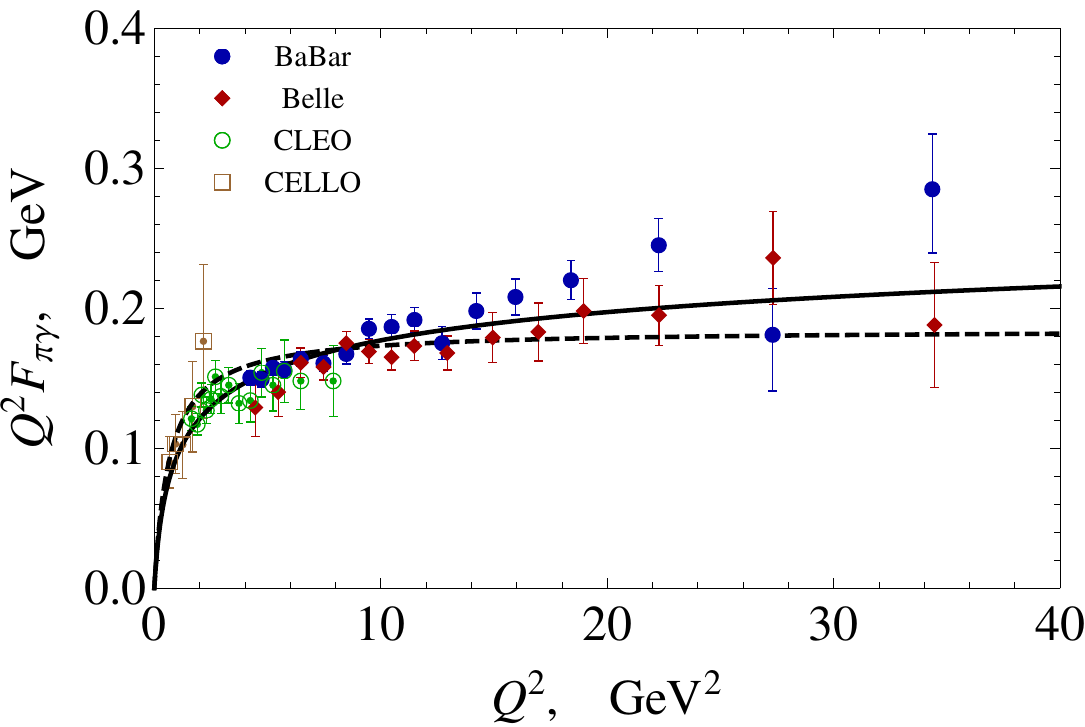}}
\caption{Pion TFF (\ref{f3m}) (dashed curve) and (\ref{corr3F}) (solid curve) compared with experimental data.}
\label{fig:1}
\end{figure}

The TFF (\ref{f3m}) matches well the experimental data  of \textit{CELLO} \cite{Behrend:1990sr}, \textit{CLEO} \cite{Gronberg:1997fj} and \textit{Belle} \cite{Uehara:2012ag} collaborations, while the data of \textit{BaBar} Collaboration \cite{Aubert:2009mc} are described much worse (see the dashed curve in Fig.\ref{fig:1}).

If the \textit{BaBar} data are correct, they strongly indicate the existence of the corrections of dimension 2 (which is absent in the local OPE) to the pion TFF. Indeed, while the full integral in the ASR has no corrections, the spectral density $A_{3}^{(3)}(s,Q^{2})$ can have corrections and therefore the pion  and continuum contributions acquire correction also (still, they should exactly compensate each other). As we mentioned above, $\alpha_s$ corrections to $A_{3}^{(3)}(s,Q^{2})$  (dimension 0) are zero; higher $\alpha_s$ corrections  are small enough to describe such a sturdy growth demonstrated by the \textit{BaBar} data;  the 4th- and higher-dimensional corrections  decrease rapidly in $Q^2$ and are also insufficient to support the \textit{BaBar} data trend.
The form of this  correction is not known (the  origins of such a correction are essentially nonperturbative). Nevertheless, we can propose \cite{Klopot:2012hd} the form of it, relying on  the general properties of the ASR. Namely, the correction should vanish at $s_3\to \infty$ (the continuum contribution vanishes), at  $s_3\to 0$ (the full integral has no corrections), at  $Q^2\to \infty$ (the perturbative theory works at large $Q^2$)  and at $Q^2\to 0$ (anomaly perfectly describes pion decay width).  If the correction contains rational functions and logarithms of $Q^2$, the simplest form of such a correction results in a pion TFF

\begin{align} \label{corr3F}
F_{\pi\gamma}(Q^2) = \frac{1}{2\sqrt{2}\pi^2f_{\pi}}\frac{s_3}{s_3+Q^2}\Bigl [1+\frac{\lambda Q^2}{s_3+Q^2}&(\ln{\frac{Q^2}{s_3}}+\sigma)\Bigr ],
\end{align}  
where $\lambda$ and $\sigma$ are dimensionless parameters. Fitting the Eq. (\ref{corr3F}) to the combined data of
\textit{CELLO} \cite{Behrend:1990sr}, \textit{CLEO} \cite{Gronberg:1997fj}, \textit{BaBar} \cite{Aubert:2009mc} and \textit{Belle} \cite{Uehara:2012ag} collaborations, one obtains the parameters $\lambda=0.12,\sigma=-2.50$ with $\chi^2/d.o.f.=0.91, d.o.f.=50$ (d.o.f.$=$number of degrees of freedom), see the solid line in Fig.\ref{fig:1}. For the dashed curve in Fig.\ref{fig:1} (without such a correction) one gets $\chi^2/d.o.f.=1.86, d.o.f.=52$. In conclusion, in the present  experimental status, the contribution of the operator of dimension 2 to the spectral density  $A_{3}^{(3)}(s,Q^{2})$ seems  preferable.

\section {Octet channel of ASR, $\eta, \eta'$ TFFs and quark mass effect}
In the octet channel of the ASR (\ref{asr}), (\ref{qhd}) the lowest hadron contributions are given by the $\eta$ and $\eta'$ mesons. The $\eta'$ meson  should be taken into account explicitly along with the $\eta$ meson, at least in the chiral limit approximation, as it gives a significant contribution to the octet channel of ASR due to the mixing, but cannot be covered by the massless one-loop form of $A_3^{(8)}$ in the limit $Q^2\to 0$.

Let us consider the "$\eta+\eta'+continuum$" model similarly to the "$\pi^0+continuum$" model in the isovector channel.  Introducing the continuum threshold in the octet channel $s_8$, using the one-loop expression for the spectral  density $A_3^{(8)}=\frac{1}{\sqrt{6}}(A_3^{(u)}+A_3^{(d)}-2A_3^{(s)})$, neglecting the $u$- and $d$- quark masses but taking into account $s$- quark mass $m_s$, one obtains from (\ref{asr}), (\ref{qhd}), (\ref{a3}):
\begin{align} \label{asr8m}
f_{\eta}^8 F_{\eta\gamma}(Q^2) +f_{\eta'}^8F_{\eta'\gamma}(Q^2)=\frac{1}{2\sqrt{6}\pi^2}\frac{s_8}{s_8+Q^2}\times \nonumber\\ \Bigl [1+\frac{4m_s^2}{3s_8}(\frac{2}{R_8^{(s)}+1}+\ln\frac{1+R_8^{(s)}}{1-R_8^{(s)}})\Bigr ],
\end{align}
where $R^{(s)}_8=\sqrt{1-4m_s^2/s_8}$. The mass term (with the opposite sign) in this equation corresponds to the higher pseudoscalar contributions in the limit $Q^2=0$, so their total contribution in this limit is negative.

 The large-$Q^2$ limit of (\ref{asr8m}) and the pQCD  predicted expression for the $\eta, \eta'$ TFFs gives $s_8$:  
\begin{equation} \label{asr8inf}
s_8=4\pi^2((f_\eta^8)^2+(f_{\eta'}^8)^2+ 2\sqrt{2} [ f_\eta^8 f_{\eta}^0+ f_{\eta'}^8 f_{\eta'}^0]).
\end{equation}

Eqs. (\ref{asr8m}) and (\ref{asr8inf}) directly relate the $\eta$, $\eta'$ TFFs with their decay constants at arbitrary $Q^2$, including $Q^2=0$, where the TFFs are expressed in terms of $2\gamma$ decay widths. 

The massless approximation of (\ref{asr8m}) ($m_s=0$) was analyzed in \cite{Klopot:2012hd,Klopot:2011qq}.
It was shown \cite{Klopot:2011qq}, that this approximation fits well the BaBar data \cite{BABAR:2011ad} with different sets of decay constants and provides an anomaly-based test for the mixing parameters. Also, it allowed to extract the decay constants in the different mixing schemes, as well as mixing scheme independently, using the experimental data \cite{Klopot:2012hd} (another recent determination of the $\eta$-$\eta'$ mixing parameters from the data on TFFs can be found in \cite{Escribano:2013kba}). We have also analyzed \cite{Klopot:2012hd,Klopot:2011ai} a possibility of the dimension 2 log-like correction, similar to the pion case (\ref{corr3F}). It appeared, that this kind of correction is not excluded by the current experimental data, but there is no a strong need for such a correction as well (unlike the isovector (pion) case).

Now, taking into account both mass (the second term in square brackets) and possible dimension 2 (the third term in square brackets) corrections, the ASR in the octet channel leads to


\begin{align} \label{asr8m-corr}
f_{\eta}^8 F_{\eta\gamma}(Q^2) +f_{\eta'}^8F_{\eta'\gamma}(Q^2)=  \frac{1}{2\sqrt{6}\pi^2}\frac{s_8}{s_8+Q^2}\times\nonumber \\ \Bigl [1+\frac{4m_s^2}{3s_8}(\frac{2}{R_8^{(s)}+1}+\ln\frac{1+R_8^{(s)}}{1-R_8^{(s)}})+\frac{\lambda Q^2}{s_8+Q^2}(\ln{\frac{Q^2}{s_8}}+\sigma)\Bigr ].
\end{align}

Note, that the quark mass term (the second one in the rhs of (\ref{asr8m-corr})) cannot provide a pion-like $\log Q^2$ growth, and the source of such behavior should be the different (the third term in the rhs of (\ref{asr8m-corr})) correction to the spectral density .   
\begin{figure}
\centerline{
\includegraphics[width=0.5\textwidth]{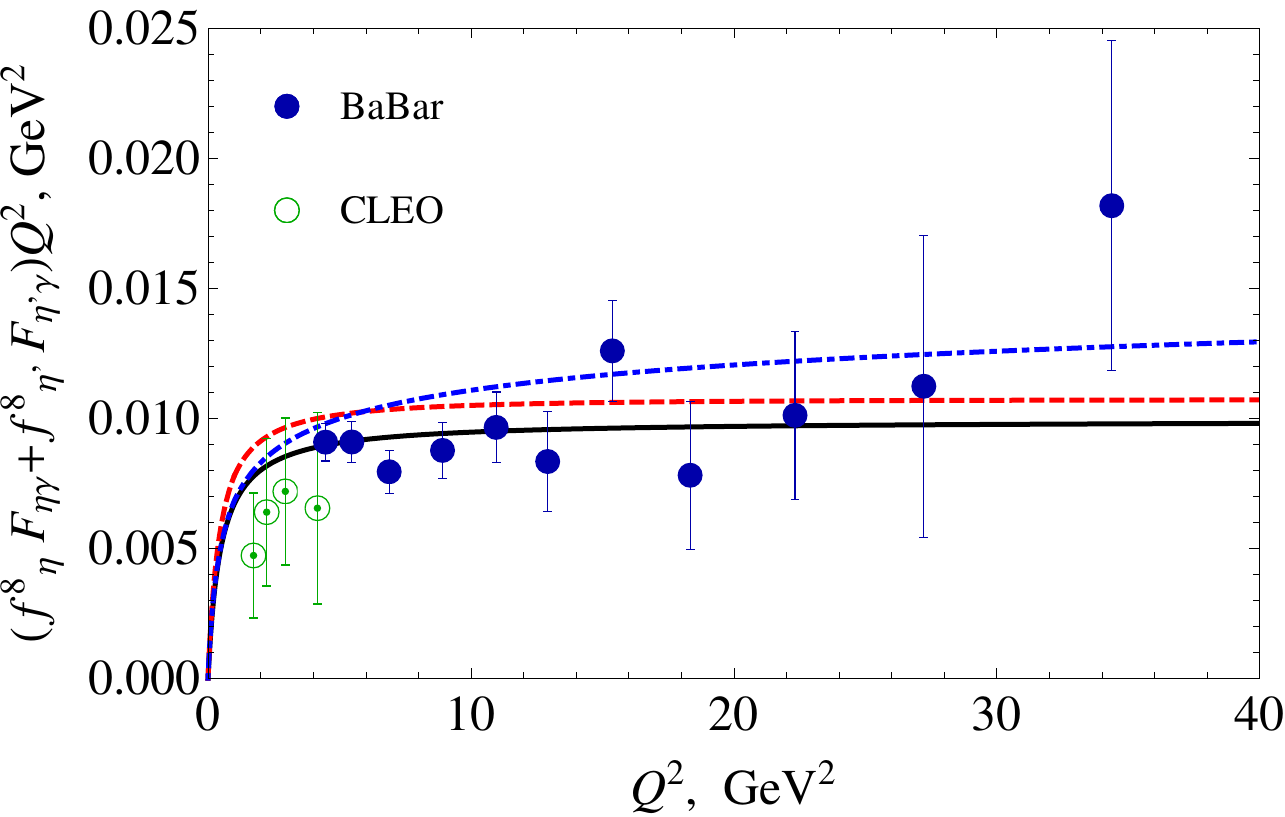}}
\caption{ASR in the octet channel (Eqs. (\ref{asr8m}), (\ref{asr8m-corr})) compared with experimental data: Solid black line -- $m_s=0$ limit of (\ref{asr8m}); dashed red line -- Eq. (\ref{asr8m}); dot-dashed blue line -- Eq. (\ref{asr8m-corr}) (see the text for details).}
\label{fig:2}
\end{figure}

In Fig.\ref{fig:2} we compare the different limits of Eqs. (\ref{asr8m}), (\ref{asr8m-corr})   with the experimental data \cite{BABAR:2011ad,Gronberg:1997fj}. The decay constants $(f_{\eta}^8,f_{\eta'}^8,f_{\eta}^0,f_{\eta'}^0)=(1.17,-0.46,0.19,1.15)f_\pi$, which correspond to the quark-flavor mixing scheme parameters $f_q=1.07f_\pi, f_s=1.34f_\pi, \phi=39.3^\circ$, are taken from \cite{Feldmann:1998vh}. 
Remarkably,  although the \textit{BaBar} data \cite{BABAR:2011ad} do not manifest a strong growing trend for the $\eta$ and $\eta'$ meson TFFs multiplied by $Q^2$ (they rather exhibit different trends at large $Q^2$), but the octet channel combination of the  TFFs (i.e. the lhs of (\ref{asr8m-corr})) multiplied by $Q^2$, does show a slight growing trend caused by the opposite signs of $f_\eta^8$ and $f_{\eta'}^8$.
The massless  ($m_s=0$) and massive ($m_s=95$ MeV at $Q^2=4$ GeV$^2$, and supposed to be frozen at $Q^2<1$ GeV$^2$) cases of Eq. (\ref{asr8m})  are denoted by a solid and dashed curves respectively in Fig.\ref{fig:2}. One can see, that although the mass contribution is not negligible ($\sim 10 \%$ of the main term), both curves fit the data decently within the experimental errors (which are quite large). We can conclude, that the massless approximation is quite reasonable.
Finally, taking into account the possible contribution of the dimension 2 correction, consider the Eq. (\ref{asr8m-corr}). Employing the parameters extracted earlier from the isovector channel, $\lambda=0.12, \sigma=-2.50$, one gets the blue dot-dashed curve in Fig.\ref{fig:2}. 

Our previous conclusion about this kind of correction  remains valid in the "massive" case also:  the octet channel of the ASR can accommodate the dimension 2 correction, although there is no such a strong demand for it as in the isovector channel.

This work is supported by  RFBR, research grants 12-02-00613a (Y.K. and O.T.) and 12-02-00284a (A.O.).

\label{}




\nocite{*}
\bibliographystyle{elsarticle-num}
\bibliography{martin}

\begin{thebibliography}{00}

\bibitem{Aubert:2009mc} 
  B.~Aubert {\it et al.}  [BaBar Collaboration],
  Phys.\ Rev.\ D {\bf 80}, 052002 (2009)
 
\bibitem{Uehara:2012ag} 
  S.~Uehara {\it et al.}  [Belle Collaboration],
  Phys.\ Rev.\ D {\bf 86}, 092007 (2012)

\bibitem{BABAR:2011ad} 
  P.~del Amo Sanchez {\it et al.}  [BaBar Collaboration],
  Phys.\ Rev.\ D {\bf 84}, 052001 (2011)
 
 \bibitem{Lepage:1980fj} 
   G.~P.~Lepage and S.~J.~Brodsky,
   Phys.\ Rev.\ D {\bf 22}, 2157 (1980).

   
\bibitem{Stefanis:2012yw} 
  N.~G.~Stefanis, A.~P.~Bakulev, S.~V.~Mikhailov and A.~V.~Pimikov,
  Phys.\ Rev.\ D {\bf 87}, 094025 (2013),

  A.~P.~Bakulev, S.~V.~Mikhailov, A.~V.~Pimikov and N.~G.~Stefanis,
  Phys.\ Rev.\ D {\bf 86}, 031501 (2012);
  Acta Phys.\ Polon.\ Supp.\  {\bf 6}, 137 (2013)

\bibitem{Agaev:2010aq} 
  S.~S.~Agaev, V.~M.~Braun, N.~Offen and F.~A.~Porkert,
  Phys.\ Rev.\ D {\bf 83}, 054020 (2011)


\bibitem{Klopot:2010ke} 
  Y.~N.~Klopot, A.~G.~Oganesian and O.~V.~Teryaev,
  Phys.\ Lett.\ B {\bf 695}, 130 (2011)

\bibitem{Klopot:2012hd} 
  Y.~Klopot, A.~Oganesian and O.~Teryaev,
  Phys.\ Rev.\ D {\bf 87}, 036013 (2013)
 
 \bibitem{Klopot:2011qq} 
   Y.~N.~Klopot, A.~G.~Oganesian and O.~V.~Teryaev,
   Phys.\ Rev.\ D {\bf 84}, 051901 (2011)
   
\bibitem{Klopot:2011ai} 
  Y.~Klopot, A.~Oganesian and O.~Teryaev,
  JETP Lett.\  {\bf 94}, 729 (2011)
   
  
\bibitem{Bell:1969ts}
  J.~S.~Bell, R.~Jackiw,
  Nuovo Cim.\  {\bf A60}, 47-61 (1969).
  S.~L.~Adler,
  Phys.\ Rev.\  {\bf 177}, 2426-2438 (1969).

\bibitem{Dolgov:1971ri}
  A.~D.~Dolgov, V.~I.~Zakharov,
  Nucl.\ Phys.\  {\bf B27}, 525-540 (1971).

\bibitem{Horejsi:1985qu} 
  J.~Horejsi,
  Phys.\ Rev.\ D {\bf 32}, 1029 (1985).
    
\bibitem{Horejsi:1994aj}
  J.~Horejsi, O.~Teryaev,
  Z.\ Phys.\  {\bf C65}, 691-696 (1995).

\bibitem{Melikhov:2012qp} 
  D.~Melikhov and B.~Stech,
  Phys.\ Lett.\ B {\bf 718}, 488 (2012)


\bibitem{'tHooft:1980xb}
  G.~'t Hooft in  
    \emph{``Recent developments in gauge theories''} ed. by G. 't Hooft \emph{et al.}, Plenum Press, New York, 1980.


\bibitem{Brodsky:1981rp} 
  S.~J.~Brodsky and G.~P.~Lepage,
  Phys.\ Rev.\ D {\bf 24}, 1808 (1981).

\bibitem{Radyushkin:1995pj} 
  A.~V.~Radyushkin,
  Acta Phys.\ Polon.\ B {\bf 26}, 2067 (1995)

\bibitem{Jegerlehner:2005fs} 
  F.~Jegerlehner and O.~V.~Tarasov,
  Phys.\ Lett.\ B {\bf 639}, 299 (2006)
 
\bibitem{Behrend:1990sr} 
  H.~J.~Behrend {\it et al.}  [CELLO Collaboration],
  Z.\ Phys.\ C {\bf 49}, 401 (1991).
  
  \bibitem{Gronberg:1997fj} 
    J.~Gronberg {\it et al.}  [CLEO Collaboration],
    Phys.\ Rev.\ D {\bf 57}, 33 (1998)

\bibitem{Escribano:2013kba} 
  R.~Escribano, P.~Masjuan and P.~Sanchez-Puertas,
  arXiv:1307.2061 [hep-ph].

\bibitem{Feldmann:1998vh} 
  T.~Feldmann, P.~Kroll and B.~Stech,
  Phys.\ Rev.\ D {\bf 58}, 114006 (1998)
  
 \end{thebibliography}


\end{document}